\documentclass[format=sigconf, screen=true, review=false, anonymous=false, timestamp=false]{acmart}

\usepackage{booktabs} 
\usepackage[utf8]{inputenc}

\usepackage{balance}

\setcopyright{none}

\usepackage{wrapfig}

\usepackage{lipsum}

\acmConference{}
\acmBooktitle{}
\acmVolume{}
\acmNumber{}
\acmArticle{}
\acmYear{}
\acmMonth{}
\acmArticleSeq{}
\acmSubmissionID{}
\acmPrice{}
\acmISBN{}
\acmDOI{}

\editor{}

\renewcommand\footnotetextcopyrightpermission[1]{} 

\begin{document}
\title{The diveXplore System at the Video Browser Showdown 2018 -- Final Notes}
\renewcommand{\shorttitle}{More Details About the diveXplore System}

\author{Klaus Schoeffmann, Bernd M{\"u}nzer, J\"{u}rgen Primus, Andreas Leibetseder}
\affiliation{%
  \institution{Institute of Information Technology}
%
  \institution{Alpen-Adria University, 9020 Klagenfurt Austria}
}
\email{{mprimus|bernd|aleibets|ks}@itec.aau.at}

\renewcommand{\shortauthors}{K. Schoeffmann et al.}

\begin{abstract}
This short paper provides further details of the \textit{diveXplore} system (formerly known as CoViSS), which has been used by team ITEC1 for the Video Browser Showdown \cite{SchoffmannBailer2012} (VBS) 2018. In particular, it gives a short overview of search features and some details of final system changes, not included in the corresponding VBS2018 paper \cite{primusvbs2018}, as well as a basic analysis of how the system has been used for VBS2018 (from a user perspective).
\end{abstract}


\maketitle

\section{System Overview}
\begin{figure*}[htbp!]
\center
\includegraphics[width=0.42\textwidth]{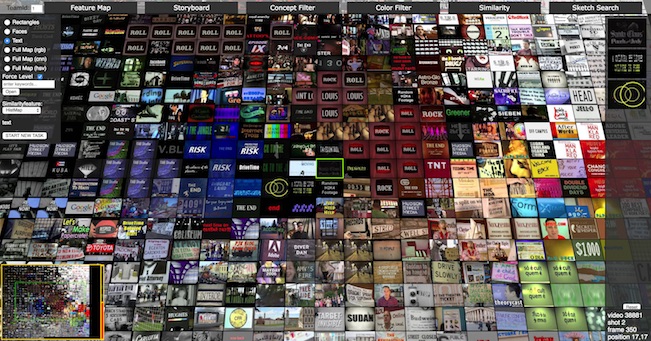}
\includegraphics[width=0.42\textwidth]{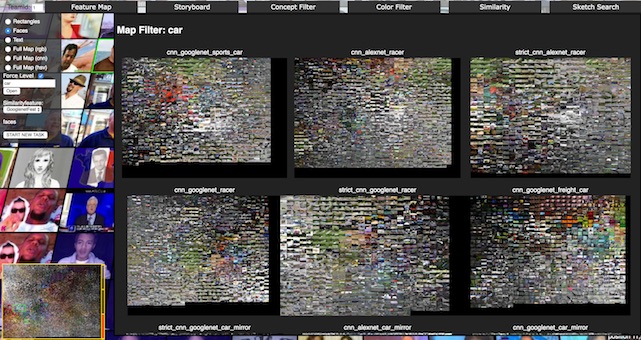}
\includegraphics[width=0.42\textwidth]{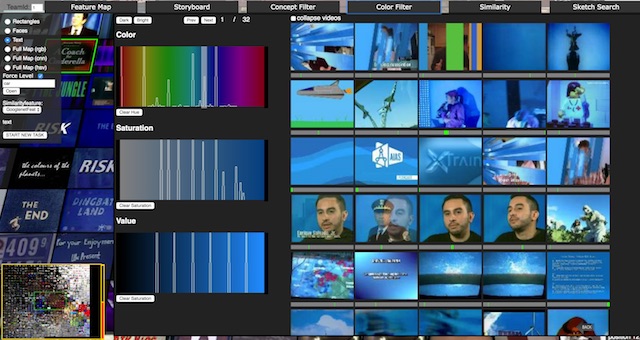}
\includegraphics[width=0.42\textwidth]{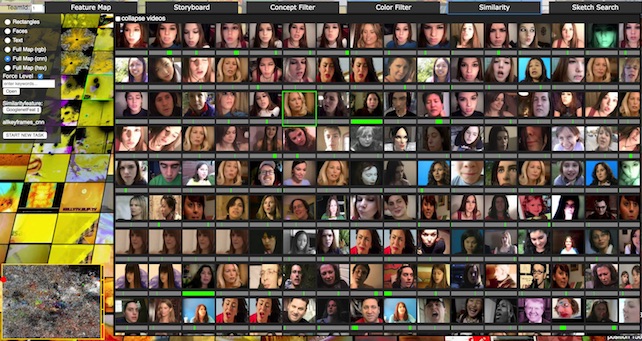}
\caption{(1) browsing feature maps of visually or semantically coherent keyframes (top-left), (2) text search for features maps (top-right), (3) color filtering/browsing (bottom-left), and (4) similarity search based on example keyframes (bottom-right).}
\label{fig:divexplore}
\end{figure*}

The \textit{distributed interactive video exploration} (diveXplore) system has been developed at Klagenfurt University for several years (and used for VBS2017 \cite{ITECUU2017} and VBS2018 \cite{primusvbs2018}) with the idea to provide a very flexible set of content retrieval features that fit many different search scenarios. Therefore, it provides a number of different components that can be used individually or in combination (see also Figure~\ref{fig:divexplore}): 

\begin{itemize}
\item browsing \textit{full feature maps} of the entire dataset  (arranged by similar color or similar classification results in CNNs).

\item browsing \textit{pre-filtered feature maps} (e.g., keyframes containing \textit{faces}, \textit{texts}, or some other semantic concept) -- for this the interface also provides a textual search feature, since there could be several hundreds of maps. 

\item a \textit{storyboard/shot view} of every video, as well as a \textit{video player} that reveals the shot structure and provides different playback speeds.

\item a \textit{color filter} to filter keyframes based on dominant colors.

\item a \textit{textual concept search} feature to filter keyframes containing specific visual concepts. 

\item a \textit{similarity search} feature for query-by-example with a selected keyframe. 

\item a \textit{color-sketch} feature (more details in \cite{leibetsedervbs18}).

\item a \textit{collaboration feature} in case the interface is used by several users, which allows to see each other's position in a map, to send hints about shots, etc. The diveXplore system also provides an additional \textit{SpectatorView} that shows the current state of all collaborators (more details in \cite{primusvbs2018}).
\end{itemize}

\begin{table*}[htbp!]
\centering
\begin{tabular}{l|c|c|c|}
\cline{2-4}
                                       & \textbf{KIS visual}                                                                                                                                 & \textbf{KIS textual}                                            & \textbf{AVS}                                                                                                                                                                                             \\ \hline
\multicolumn{1}{|l|}{\textbf{Experts}} & \begin{tabular}[c]{@{}c@{}}concept search or sketch search\\ (typically in combination\\ with shot filtering and  \\ video inspection)\end{tabular} & \begin{tabular}[c]{@{}c@{}}mostly\\ concept search\end{tabular} & \begin{tabular}[c]{@{}c@{}}a combination of \\ concept search, map search,  \\ map browsing, and similarity search\\ (typically in combination with \\ shot filtering and video inspection)\end{tabular} \\ \hline
\multicolumn{1}{|l|}{\textbf{Novices}} & \begin{tabular}[c]{@{}c@{}}mostly \\ concept search\end{tabular}                                                                                    & -                                                               & \begin{tabular}[c]{@{}c@{}}mostly concept search\\ (often with shot filtering), \\ sometimes map search \\ and browsing\end{tabular}                                                                     \\ \hline
\end{tabular}
\caption{Search Feature Usage of diveXplore at VBS2018}
\label{tab:usage}
\end{table*}

\section{Final System Changes}

The system described in our VBS2018 paper \cite{primusvbs2018} has been further developed and improved after paper submission. Therefore, we describe these changes here in more detail.

First of all, we significantly improved our feature map browsing component in several ways: (i) all feature maps are now arranged by using a self-organizing map algorithm with some similarity feature (e.g., color or semantic concept), (ii) we provide pre-filtered feature maps that contain only keyframes of a specific concept (e.g., faces or texts -- see top-left in Figure~\ref{fig:divexplore}). This includes maps created for concepts that have been recognized in a large enough volume (at least 576 corresponding keyframes) in the dataset by different CNNs. Since this resulted in more than 1200 feature maps, we have also implemented a \textit{textual map search} feature (Figure~\ref{fig:divexplore}).   

Next, the \textit{storyboard} used for VBS2017 \cite{ITECUU2017} -- that consisted of small and uniformly sampled thumbnails -- has been replaced by a \textit{shot view} of each video. Similarly to the previous version, it allows to browse over different videos of the dataset quite quickly and easily (and to jump to one specific video). 

We introduced a new \textit{similarity} tab in our interface that always displays the results of the previous similarity search. This facilitates jumping between different components of our interface (e.g., looking into the shot view of a video and then going back to the similarity search results). In addition to that we implemented a history feature that allows going back to previously retrieved search results.  

Finally, we disabled the \textit{web-example} search component (see \cite{primusvbs2018}) in our system, because we found it hard to use for the IACC.3 dataset \cite{2017trecvidawad}, due to its low visual quality. Instead we integrated a color-sketch feature in collaboration with our second team \cite{leibetsedervbs18}.   


\section{System Usage at VBS2018}

The usage of diveXplore at VBS2018 differs heavily for the two different user groups (\textit{experts} and \textit{novices}), as indicated in Table~\ref{tab:usage}. While novices mostly used textual concept search, the experts utilized many more features of the system and followed a more diverse search strategy. However, due to the different type of query presentation (visual KIS queries are presented as video clips, while textual KIS and AVS tasks are described by text), the strategy for completing visual tasks is much more visually-oriented (hence, users closely inspected the video and sometimes used sketches). For textual KIS the experts mostly employed simple textual concept search, since no or merely a few clues about the visual representation of the scene were provided. For AVS, however, due to the fact that multiple correct submissions are possible (and their correctness may be immediately confirmed by the VBS Server during the task), users tend to use additional features that could further improve the performance (i.e., allow them to find more instances more easily/quickly), such as map search (followed by browsing filtered maps) and similarity search based on an already found instance \cite{kletzMMM18}.

\section{Conclusions}

The usage experience of our system gained at VBS2018 clearly shows that the strategy to build a flexible system with many different retrieval options is the right one. Different type of queries need different search features and interactions means, this is also evident from the many different interfaces already proposed in the literature for video search and interaction \cite{Schoeffmann2015Survey}. We will continue to improve our interactive video retrieval system diveXplore and try to further improve our performance, which was quite solid in the past two years (2nd place at VBS2017 and VBS2018).

\bibliographystyle{ACM-Reference-Format}
\balance

\bibliography{bib}

\end{document}